\documentstyle[twocolumn,aps,prb,epsfig,floats]{revtex}
\draft

\begin{document}
\twocolumn[\hsize\textwidth\columnwidth\hsize\csname@twocolumnfalse\endcsname

\title{A hierarchical approach for computing spin glass ground states}
\author{J\'er\^ome Houdayer$^{(1)}$ and Olivier C. Martin$^{(2)}$}

\address{
1) Institut f\"ur Physik, \\
Mainz, Germany \\
2) Laboratoire de Physique Th\'eorique et Mod\`eles Statistiques, \\
b\^at. 100, Universit\'e Paris-Sud, F--91405 Orsay, France}

\date{\today}
\maketitle

\begin{abstract}
We describe a numerical algorithm for computing spin glass
ground states with a high level of reliability. The method
uses a population based search and applies
optimization on multiple scales. Benchmarks
are given leading to estimates of the
performance on large lattices.
\end{abstract}


\twocolumn]\narrowtext

\section{Introduction}
\label{sect_intro}
Discrete optimization plays a central role in many engineering
problems such as scheduling and electronic circuit design, but it is also
important in fundamental science. One major open problem
there concerns the nature of the energy landscape of
optimization problems with quenched disorder.
It is generally agreed that
these energy landscapes are rugged, but are they
self-similar as predicted by the scaling theories
of spin glasses~\cite{BrayMoore85,FisherHuse86}?
Despite many years of controversy, this issue is still
unsettled. In order to make progress, it is
useful to be able to compute the
ground states of these disordered systems~\cite{AlavaDuxbury01}.
Indeed, by applying sensitivity analysis, that is
by considering how the ground state changes 
when the parameters specifying the optimization problem vary, one
can probe the energy landscape and measure its
scaling exponents. For some types of optimization problems,
finding ground states can be achieved with algorithms whose
CPU time grows polynomially with the problem
size (the size is the number of discrete variables in
the problem). However, in many other interesting cases,
finding the ground state is an NP-hard problem 
so {\it efficient} algorithms are particularly called for.
In this work, we present a method for computing
ground states heuristically; even though our algorithm
is not guaranteed to provide the optimum as its output,
the ground states can be found for significantly larger
systems than with methods having such a guarantee. For instance
the ground state of a $10\times 10\times 10$ 
Edwards-Anderson spin glass 
can be computed in a few minutes on a personal computer, while
$13\times 13 \times 13$ lattices can be solved on larger computers.
Results using exact methods such as branch and bound
have been published~\cite{KlotzKobe94} but only for much smaller
sizes ($4 \times 4 \times 4$).

In the next section, we 
give the general framework in which we work.
Section~\ref{sect_components}
introduces the main features of our algorithm which 
embeds both a local search and renormalization into a
genetic algorithm; then section~\ref{sect_architecture}
describes the over-all algorithm architecture. Finally,
in section~\ref{sect_behavior}, we
explain how the algorithm behaves in practice on 
benchmark problems.

\section{Graphs and Hamiltonians}
\label{sect_graphs}
We consider an Ising spin glass, defined on an arbitrary 
non-oriented graph $G$; 
the Hamiltonian or energy function we seek to minimize is
\begin{equation}
H = - \sum_{\{i,j\} \in E} J_{ij} S_i S_j - \sum_{i \in V} h_i S_i \ .
\label{eq_hamiltonian}
\end{equation}
The ``spins'' $S_i$ are of the Ising
type, $S_i = \pm 1$, and lie on the
vertices or ``sites'' of $G$. $N$ is the {\it size} of our system, that
is the number of these vertices.
$E$ is the set of edges of $G$; each edge connects
two vertices. Finally, the
$J_{ij}$ and the $h_i$ can be any real numbers;
the $J_{ij}$ lie on the edges of $G$ while
the $h_i$ lie on its vertices. From here on, 
we call an {\it instance} the
specification of all the parameters of the Hamiltonian,
that is the specification of the graph $G$ and of all the
parameters $J_{ij}$
and $h_i$. Similarly, we call a (spin) {\it configuration} the 
assignment of the values of the $S_i$ (for all $i$,
$1 \le i \le N$).

The Hamiltonian in equation~\ref{eq_hamiltonian} can be used
to represent an arbitrary spin glass, be it on a lattice like the
Edwards-Anderson~(EA) model~\cite{EdwardsAnderson75}, or on a 
random graph as for diluted mean field
spin glasses. It can also
be used to represent a random field
Ising model (RFIM) with or without disorder
in the bond strengths. The algorithm we present in this work does not
take advantage of any structure in $G$ or in the
parameters defining the Hamiltonian; because of this,
it cannot be expected to be competitive in the
special cases where finding the ground state is
a polynomial problem. (Note that $2$-dimensional
spin glasses and the RFIM fall into this
class.) Nevertheless, in the other
cases we have found the algorithm to be very effective. It is
possible that improvements could be realized
by taking advantage of additional structure in
$H$, but we shall not investigate that issue here. 

\section{Components of the algorithm}
\label{sect_components}
The algorithm we present is a genetic algorithm:
we evolve a population of configurations from one
generation to the next. The key elements of our
approach are: 
(i) the use of an embedded local search;
(ii) the incorporation of a renormalization 
procedure among parents in the population which allows
one to consider multiple length scales;
(iii) the use of recursion. Because of these features, 
we talk of a ``Genetic Renormalization Algorithm'' (GRA).

A previous GRA algorithm was described
in reference~\cite{HoudayerMartin99b}. For the readers aware of that
work, our new approach has a different
structure for the recursive calls, leading to 
an exponential speed-up in $N$ compared to the
older method. Before
describing our new algorithm, we first go over its main
components.

\subsection{Local search}
\label{subsect_local_search}
Let us first define a few terms. Consider a configuration and one of
its spins $S_i$.
We call {\it gain} of that spin the
decrease in the configuration's energy when $S_i$ is flipped. (The
gain is thus minus the change in the total energy.)
A spin is said {\it stable} (resp. {\it
unstable}) if its gain is negative (resp. positive). 
Spin $S_i$ is said to be more unstable than 
spin $S_j$ if $gain(S_i) > gain(S_j)$.

It is straightforward to extend these single spin definitions
to {\it sets} of spins. We shall focus on the case of
clusters; a {\it cluster} is a set of spins (vertices)
which is connected (using the edges
of $G$). Then, extending the previous definitions, the gain
of a cluster is just minus the change of the total energy when all the
spins of the cluster are flipped. Similarly, a cluster can be stable
or unstable, etc...

A local search is a procedure that attempts to lower the
energy of a configuration by repeatedly changing a few
variables at a time, accepting only improving changes. For our system, 
this amounts to flipping sets of spins with positive gain until no
more favorable sets are found. Local search covers many different
methods because there is much freedom in the way one searches for
and selects these sets. The method
we have developped for our GRA is inspired from the
Kernighan-Lin~\cite{KernighanLin70} algorithm. In 
that class of local search algorithms, the number of variables that is
changed at a time is not set before-hand and in 
practice can be quite large. For our local search,
we force the set of spins that will be flipped to be connected,
that is we restrict ourselves to clusters.
Indeed, if there is a set of spins with positive gain,
at least one connected component of that set has a positive gain. 

Our search for a ``good'' cluster proceeds as follows.
First we choose a starting
spin; it defines the initial cluster. Second, we successively add 
new and promising spins to the current cluster, maintaining the
connectivity property. During this growth process, the gain of
the cluster can go positive or negative,
up or down. Third, we stop growing the cluster when things
no longer look promising. Finally, we consider all the gains 
generated during the cluster's growth
and select the largest one. If that gain is strictly positive,
we flip the corresponding cluster, generating an improved
configuration. These steps correspond to one
pass of the local search. We perform multiple passes until the search
for an improving cluster fails; then 
a local minimum of $H$ has been reached and the local search is 
finished. 

Naturally, the description we have given of our local search is rather
schematic; one has to implement in the code how
to choose the starting spin, what is a promising spin, etc...
The reader interested in these details will find them in
Appendix 1. Other types of local searches could be used
instead; our choice is motivated by a trade-off 
between speed and quality.

\subsection{Renormalization}
\label{subsect_renormalization}
Given a spin glass Hamiltonian for $N$ spins and 
at least two spin configurations,
there is a natural definition of a block spin; from that,
one can extract an exact renormalized spin glass Hamiltonian
associated with a system having in general $N' < N$ spins.
Such a renormalization procedure was first proposed by Kawashima and
Suzuki~\cite{KawashimaSuzuki92}; to keep this paper self-contained,
we describe this method again.

Let equation~\ref{eq_hamiltonian} be the Hamiltonian
of the system and $G$ be the associated graph. Suppose we have 
$k$ spin configurations $\{ S_i^{(1)}\}$,  
$\{S_i^{(2)}\}$, \ldots, $\{S_i^{(k)}\}$. We then
define the {\it signature} at site $i$ to be the
following vector of $\pm 1$
values:
\begin{equation}
{\vec \sigma}_i = (S_i^{(1)} S_i^{(2)}, 
S_i^{(1)} S_i^{(3)}, \ldots, S_i^{(1)} S_i^{(k)}).
\label{eq_signature}
\end{equation}
Two sites $i$ and $j$ have the same signature
if and only if the spins at those sites have the
same {\it relative} orientation (parallel or anti-parallel)
in {\it all} of the $k$ configurations.
Now we partition
the sites of $G$ according to their signature. Furthermore, given
a set of sites of identical signature, we further subdivide this set
into clusters ({\it i.e.}, the desired subsets are the connected components
of that set, where as usual connectivity is defined using the edges
of $G$).  Thus to each site $i$ corresponds a maximal connected
cluster $A(i)$ of sites; then for any site $j$,
\begin{equation}
A(j) = A(i) \Rightarrow {\vec \sigma}_j = {\vec \sigma}_i.
\label{eq_same_signature}
\end{equation}
For each cluster $A$, we introduce the ``block-spin''
$S_A$ to be $+1$ if all the spins in $A$ are parallel to
those of the first configuration, and $-1$ if they are all
anti-parallel. Thus for configuration $1$ we have
$S_A = 1$ for all $A$ while for the $n$th configuration
($1 \le n \le k$) we have
\begin{equation}
S_A^{(n)}= S_i^{(1)} S_i^{(n)}, \mbox{ for any } i \in A.
\label{eq_spin_match}
\end{equation}
as can be seen from 
equations~\ref{eq_signature} and~\ref{eq_same_signature}.

\begin{figure}
\centering
\epsfig{file=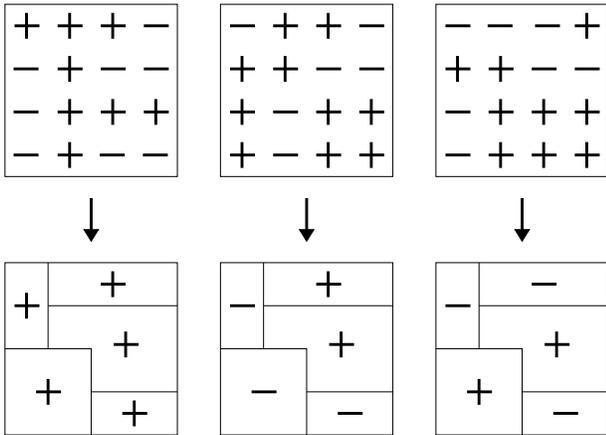,width=0.45\textwidth}
\caption{Example of a renormalization with $N=16$ and $k=3$
for nearest neighbor interactions. If the 
interactions had been long range, the top left and the bottom right clusters
would be merged since they would form a connected set of sites with the
same signature.}
\label{fig_renorm}
\end{figure}

The exact Hamiltonian for these block spins is of the form
\begin{equation}
H'[\{S_A\}] = - \sum_{(A,B) \in E'} J_{AB} S_A S_B - \sum_{A \in V'} h_A S_A,
\end{equation}
where $V'$ is the set of the $A$ clusters
and $E'$ is the set of edges of
the corresponding ``renormalized'' graph $G'$. Note that two 
blocked spins $S_A$ and
$S_B$ are neighbors in $G'$ if at least one spin of the cluster
$A$ is a neighbor (according to $G$) of at least one spin of the
cluster $B$. A simple computation gives the
couplings between the block spins:
\begin{equation}
J_{AB} = \sum_{i \in A} \sum_{j \in B} J_{ij} S_i^{(1)} S_j^{(1)} ~.
\end{equation}
Similarly, the fields are:
\begin{equation}
h_A = \sum_{i \in A} h_i S_i^{(1)}  ~.
\end{equation}
Thus we have
\begin{equation}
H[S_i^\alpha] = H'[S_A^\alpha] + H_c,
\label{eq_hamiltonian_match}
\end{equation}
where $H_c$ is a constant representing the interaction energy inside
the clusters
\begin{equation}
H_c = - \sum_{A \in V'} \sum_{i,j \in A} J_{ij} S_i^{(1)} S_j^{(1)}.
\end{equation}
In figure~\ref{fig_renorm} we give an illustration of
the construction of the clusters (block spins) generated by 
the renormalization
when using $3$ configurations. (In this example,
$G$ is a $4 \times 4$ lattice.)

The renormalization procedure can be reversed in the sense that if
you have a spin configuration for $H'$ you can recover the
corresponding one for $H$ using equation~\ref{eq_spin_match} (provided
you remember $S_i^{(1)}$ and the definition of the clusters).  We will
call this operation the {\it raising} of a configuration.

The idea is then as follows. If you have a set of configurations
for $H$, you first use the renormalization procedure to 
produce configurations $\{S_A^{(n)}\}$ associated with a smaller
number of spins (these are the block spins). Second, you improve
these configurations by a suitable optimization procedure. Finally,
you raise these improved configurations, obtaining new
configurations for the initial Hamiltonian but with
lower energies than previously.

\subsection{Population evolution}
\label{subsect_population}
The main problem encountered when searching for the ground state
via local search is that asymptotically one has
a fixed percentage error on the (extensive) ground state energy and so
the probability of reaching the ground state goes to zero
exponentially with $N$. To postpone this bad behavior one could 
improve the local search but that
is computationally costly. Furthermore,
in spin glasses (as in most difficult optimization
problems), low energy configurations may differ from the ground
state by a ``large'' number of
spins, in fact a number growing linearly with $N$.
(This is expected to happen when the overlap probability
distribution $P(q)$ is broad, signaling replica symmetry
breaking~\cite{MezardParisi87b}). When this happens,
improvements in the local search algorithm are doomed to be
ineffective. So instead we appeal to renormalization
in order to optimize on larger length scales 
while still using the {\it same} local search! As an extra bonus,
one may be able to perform optimization on all scales if the
renormalization is done recursively...

Using the local search and the renormalization together
requires working with a {\it population} of configurations.
Our method is thus a generalized genetic algorithm; we evolve a
population of configurations whose energies we try
to minimize. To do this, we repeatedly choose subsets of the population
(the {\it parents}) to which we apply the renormalization
procedure. For each such subset, we then
obtain a smaller spin glass instance and a corresponding set of
configurations. We optimize (see next section) these configurations and
deduce new configurations for the original Hamiltonian (the {\it
children}) that have a lower energy than their parents. When all the
parents have produced enough children, the parents are replaced by the
children, giving rise to a new generation.  When no further
improvements are possible, the algorithm stops.

One of the essential features of our approach is that we impose
the children to be better than their parents. Let us define this
notion mathematically as follows.
We say that a population $P'$ is {\it more optimized} than a population
$P$ if there is a mapping $f$ from $P$ to $P'$ such
that: (i) $f$ is {\it onto}, {\it i.e.}, the mapping
of $P$ covers all of $P'$; 
(ii) $f(C)=C' \Rightarrow H(C') \le H(C)$. Furthermore, we
say that $P'$ is strictly more optimized than $P$ if
$f$ is not one-to-one or if $H(C)> H(f(C))$ for at least
one $C$. These definitions
introduce a {\it partial ordering} relation on populations.
According to this relation, you obtain a strictly more
optimized population in the following cases: (i) you remove the
worst configuration (greatest energy); 
(ii) you remove duplicated configurations;
(iii) you improve at least one configuration; 
(iv) you replace a sub-population by a strictly more
optimized one. In our algorithm, duplicated configurations are
removed, so all the configurations are different in any
given population.

\section{Architecture of the algorithm}
\label{sect_architecture}

\subsection{Recursion: the Next\_Generation function}
\label{subsect_Next_Generation}
We can now discuss more precisely how the algorithm works and in
particular how the children are produced using a {\it recursive}
call to the renormalization procedure. The heart of this is a
function that we call 
``Next\_Generation''. We first describe its main ingredients;
details will be added later in
Appendix 2. For its input, this function takes an instance
({\it i.e.}, a Hamiltonian) and a population of configurations (hereafter
called the old generation). It outputs a new generation that
is more optimized than the old one. This function proceeds
as follows (see figure~\ref{fig_next_gen}):

\begin{enumerate}
\item If the old generation contains only one configuration, do nothing and
return that configuration (the new generation is the same as the old one).
\item Choose $k$ configurations (or parents) at random in the old generation.
If possible choose different parents each time.
\item Apply the renormalization procedure to create a renormalized
instance together with a set of $k$ renormalized configurations.
\item Apply the local search to these renormalized configurations.
\item Call Next\_Generation (recursively) on the renormalized instance and the
renormalized configurations. Raise the resulting configurations and apply
local search to them. These configurations are the children.
\item Add the children produced to the new generation.
\item Return to step 2 until all the configurations from the old generation
have been used as parents at least once.
\item Return the new generation (all the children produced).
\end{enumerate}

\begin{figure}
\centering
\epsfig{file=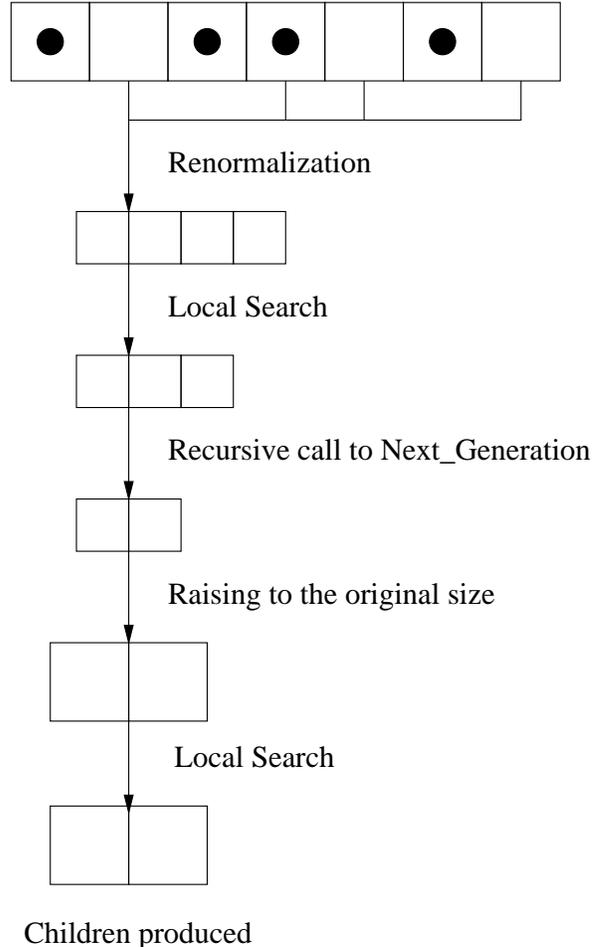,width=0.45\textwidth}
\caption{The inner steps of Next\_Generation. The dots mean that
the parent has already been used and thus should not
be re-used if possible. The steps are
repeated until all configurations have been used as parents 
({\it i.e.}, have obtained a dot).}
\label{fig_next_gen}
\end{figure}

The spirit of this function is roughly that each configuration from
the old generation gives birth to a more optimized one in the new
generation. This new configuration is essentially the old one
optimized on many scales by the local search during the recursive
calls. In the over-all genetic algorithm (see the next section), this
function is called repeatedly; thus previous improvements can
influence the ones to come.

\subsection{Layout of the algorithm}
\label{subsect_layout}
At a high level of description, 
our algorithm computes an optimized configuration 
(hopefully the ground-state) when it is
given an instance. Schematically, it proceeds
by following these $5$ steps:

\begin{enumerate}
\item Randomly generate $M$ configurations.
\item Apply the local search to each one of them.
\item Call the Next\_Generation function.
\item Return to step 3 until only one configuration is left.
\item Output this last configuration which is the result.
\end{enumerate}

There is still one detail which must be fixed. We want to be sure that
the loop terminates, and thus we want each generation to be
{\it strictly} more optimized than the previous one. We enforce
this at the level of Next\_Generation as 
explained in Appendix 2.
Given this last detail, the high-level description of the
algorithm is now complete; let us now go on and see how well
all this works in practice.

\section{Behavior of the algorithm}
\label{sect_behavior}
In the rest of this article, we restrict ourselves to 
Edwards-Anderson (EA) spin-glasses in
dimensions $2$, $3$, and $4$ with periodic boundary
conditions and nearest neighbor interactions. Furthermore,
our couplings are Gaussian and there is no magnetic field.
Thus in equation~\ref{eq_hamiltonian},
$h_i =0$ and the $J_{ij}$ are
independent random variables having a normal distribution
of zero mean and variance equal to $1$. (Generally speaking, the algorithm 
has an easier time finding ground states when there is a magnetic
field, justifying our choice of $h_i=0$.)

To give the reader some intuition about 
how the algorithm works, we can follow what happens when going
from one generation to the next; that is the object of the next
subsection. In the second subsection, we shall see
just how powerful the algorithm is by measuring the probability
with which it finds the ground state.

\subsection{Qualitative aspects}
\label{subsect_qualitative}
In our algorithm, there is only one free parameter:
the number $M$ of configurations in the first
generation of the genetic algorithm. 
(The other parameters have 
been fixed once and for all in the code which 
thus becomes a ``black-box'' routine.) 

The choice $M=1$ is
special as it prevents any renormalization; the algorithm
then reduces to applying local search to a randomly
generated configuration. How well does that work?
Let $\Delta E$ be the difference between the output 
energy and the ground state energy $E_0$. We find empirically
that $\Delta E$ is self-averaging as the linear lattice
size $L$ goes to $\infty$. Quantitatively, the relative 
error $\Delta E / E_0$ at large $L$ is $5.1\%$ in $d=2$,
$6.0\%$ in $d=3$, and $6.2\%$ in $d=4$. This may seem
large but it should be compared to the excesses of over
$20\%$ obtained when using
single spin optimization (zero temperature Metropolis).

When $M>1$, the first thing the algorithm does is apply local
search to the configurations; thus before calling Next\_Generation,
we have configurations with the previously given excess energies.
Then, as one goes from one generation
to the next, the mean energy 
of the population decreases. This is illustrated
in figure~\ref{fig_evolution} for a typical instance
with $L=12$, $d=3$ and $M=1000$. We see that this decrease is initially
quite rapid, while at the same time the population size stays fixed
at its initial value $M$. For later
generations, the mean energy decreases more slowly whereas
the population size decreases steadily. Finally, the
population size reaches $1$ and the algorithm terminates; the
last configuration is necessarily the best
configuration found throughout the whole run. The pattern shown
in the figure
is typical and arises for all lattice sizes and dimensions
we have investigated.

Naturally, the detailed evolution from
generation to generation does fluctuate from
instance to instance. Nevertheless, we find that
the generation number where the population begins
to decrease is quite insensitive to the value of
$M$ while it clearly grows for increasing $L$.
Furthermore, in the great majority of cases,
we find that the final configuration (which is the output) first appears
right after the population begins to decrease. 
Probably the most significant dependence on $M$ concerns
the total number of generations produced before termination. 
The fluctuations in that number
are larger than for other observables; also, the trend is 
towards more generations as $M\to \infty$.
But even there the dependence is not so dramatic; to give
some illustrative numbers, consider again $L=12$ for $d=3$. 
There are typically
$10$ generations for $M=100$; when $M$ increases, the number
of generations increases quite slowly, reaching about
$20$ for $M=10000$.

\begin{figure}
\centering
\epsfig{file=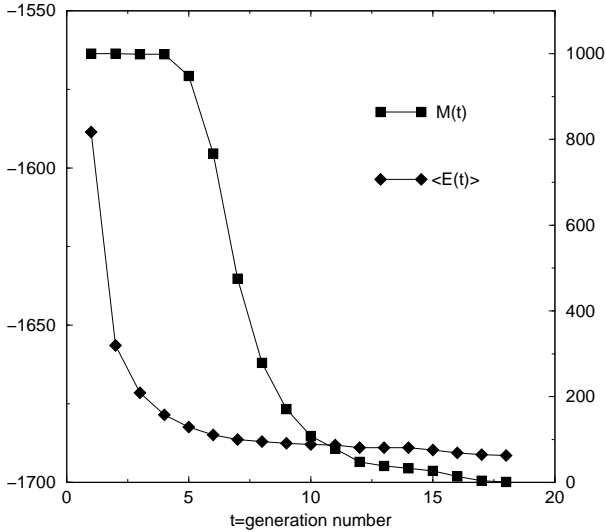,width=0.45\textwidth}
\caption{The evolution with generation number of
the mean energy (diamonds and left axis) and of the population size 
(squares and right axis).}
\label{fig_evolution}
\end{figure}

\subsection{Probability of finding the ground state}
\label{subsect_probability}
To a large extent, the power of an optimization
algorithm can be quantified by its
probability of finding the ground state and by
the CPU time required to do so. 
%
For a fixed instance and a given
number $M$, there is a certain probability $p$ of finding the ground
state (remember that our algorithm is stochastic and thus depends on a
random number generator). To measure $p$, one should know the ground 
state, but since that is not the case we proceed self-consistently.
We choose a large $M$ and we run the program many times;
if the best output (the putative ground state) is found with a high
probability (say $90\%$) we can reasonably expect that it is the ground 
state. Morevover we checked the program against a previous version
which had been also tested against an exact
algorithm~\cite{HoudayerMartin99b}.

Once we ``know'' the ground state, we can measure $p$ for different values
of $M$. Empirically we find that for large enough $M$ we have:
\begin{equation}
p(M) \approx 1 - e^{-M/a}
\label{eq_p_formula}
\end{equation}
where $a$ is a number which depends on the actual instance. This
kind of dependence on $M$ can
be motivated by a very simple argument. Let us call $q=1-p$, the
probability of not finding the ground state. If we run the program
$n$ times (with different random numbers), the probability of not
finding the optimum $n$ times is $q^n$. If we now run the program with a
population of size $n M$ instead of $M$, 
we should do at least as well and thus
$q(nM)\leq q(M)^n$. But for large $n M$, the algorithm is expected
to do no better, in which case we have the equality
$q(nM)=q(M)^n$ when $n \to \infty$; this leads directly to
equation~\ref{eq_p_formula}.

The value of $a$ depends on the instance studied.
However, for our benchmarks, we are interested
in {\it randomly} generated instances of the EA model.
In that case, we find that the fluctuations of
$a$ decrease as $L$ increases, suggesting
that $a$ is self-averaging. Table~\ref{tab_a_values} shows 
the (mean) values of
$a$ obtained from our runs.
It is not clear what the function $a(L)$ is,
but as a first guess $a$ seems to grow exponentially with $L$ (at
least in dimensions $3$ and $4$).
\begin{table}
\caption{The value of $a$ (equation~\ref{eq_p_formula}) for EA
spin glasses in dimensions $2$, $3$ and $4$. $L$ 
is the linear size of the
system. The value marked with a (?) are unreliable because it 
was not clear whether the ground states were found.}
\label{tab_a_values}
\begin{center}
\begin{tabular}{||c|c||c|c||c|c||}
\hline
\multicolumn{2}{||c||}{2D} & \multicolumn{2}{c||}{3D} &
\multicolumn{2}{c||}{4D}\\
\hline
$L$ & $a$ & $L$ & $a$ & $L$ & $a$\\
\hline
10 & 5.3 & 4 & 4.4 & 3 & 5.6\\
20 & 9.4 & 6 & 14 & 4 & 27\\
30 & 11 & 8 & 53 & 5 & 230\\
40 & 16 & 10 & 200 & 6 & 5500(?)\\
60 & 27 & 12 & 650 & & \\
80 & 43 & 14 & 2500(?)& &\\
\hline
\end{tabular}
\end{center}
\end{table}
The {\it speed} of the program appears to be roughly linear in $N$ and $M$
(for large values). On a 180MHz PC running
under Linux the coefficient of this law is
$1.2$ $10^{-4}$ sec. which means that with $M=a$ one requires 24
seconds to find the ground state of a $10^3$ EA spin-glass and 135 seconds for
a $12^3$ spin-glass.

\section{Discussion and conclusions}
\label{sect_conclusions}
As with all known algorithms, it becomes
increasingly difficult to find the true ground state
as the number of spins increases. What is important though
is that larger systems can be 
tackled~\cite{HoudayerKrzakala00} with our algorithm
than with previous methods~\cite{Pal96b,PalassiniYoung00a}.
Indeed, before the advent of GRA type algorithms, 
it was not possible to solve reliably Gaussian
EA models of sizes beyond $8 \times 8 \times 8$. (Let us 
note however that larger sizes can be tackled for the
$\pm J$ model; both Pal~\cite{Pal96a} and
Hartmann~\cite{Hartmann99} quote results for sizes up to
$14 \times 14 \times 14$.) Interestingly, if one looks
at currently competitive methods, they all rely on
genetic algorithms; it thus seems essential to use multiple
configurations (parents) to find very low energy configurations
in spin glasses. Our approach takes advantage of this
while at the same time allowing optimization on
multiple scales; we believe this is the source of our
extra performance.

There is every reason to believe that our hierarchical
approach can be useful for other optimization problems. 
Not only is the concept of a block spin natural, but
also the use of recursion is not specific to spin glasses.
In fact, we have shown previously that a GRA type
approach is effective~\cite{HoudayerMartin99b}
for the traveling salesman problem; applications
to other problems should appear soon. Naturally,
to get the best possible performance, it is necessary
to have a good local search method and to 
define intelligently the renormalization transformation
for the problem at hand. A certain amount of 
problem-specific fine tuning
is possible here. In addition, it would be useful to
investigate improvements to our GRA that are 
{\it problem-independent}.
Among these, we consider particularly promising the
possibility of 
selecting the parents in a non-random fashion in the
Next\_Generation function, and the maintaining of
diversity of the population as the generation number increases.

\vspace{1cm}
\centerline{\bf Acknowledgments} 

\bigskip
We thank K. Pal and W. Krauth for helpful comments. 
J.H. thanks the Max Planck Institute for Polymer Research
for its support during this work. The LPTMS
is an Unit\'e de Recherche de l'Universit\'e Paris~XI associ\'ee au
CNRS.  

\vspace{1cm}

\centerline{{\bf Appendix 1: Local search}}
Our local search algorithm proceeds as follows:

\begin{enumerate}
\item Choose a spin that will be the ``seed'' of the growing cluster.
This is done by taking any
of the strictly unstable spins of the current configuration;
if there are none, choose a spin at random anywhere.
Virtually flip this spin and mark it so it cannot
be flipped again at any time during the cluster's growth.
Compute the new
gains of all the other spins of this modified configuration.
\item Add to the cluster the spin with the highest gain,
be-it positive or negative with the
constraint that the cluster must remain
connected. Update the configuration and the gains
and again mark this spin so it will not be considered
for flipping during the growth of this cluster.
\item Return to step 2 unless there are no more spins to add to the cluster
or more than 20 spins have been added to the cluster since the greatest
gain has been encountered.
\item If the best cluster encountered during the growth process has a
strictly positive gain, flip it.
\item Return to step 1 unless step 4 has already failed 3 times to
find a cluster to flip, {\it i.e.}, all clusters found
had non-positive gains.
\end{enumerate}

At the end of this search, all spins have non-positive gains,
so we guarantee that the configuration is at least
1-spin-flip optimal.
One of the important features of this Kernighan-Lin-like
algorithm is that the 
size of the cluster is not limited (except by $N$). It can be very large
in practice, especially when the original configuration is random. In
order to implement this algorithm efficiently,
we use adapted data structures. First, we 
dynamically maintain a list of the unstable spins; this allows
step 1 to be done in time $O(1)$. Second, we use a dynamically maintained
heap structure to find the next best spin to add to the cluster; this
allows step 2 to be achieved in time $O(\ln K)$ where $K$ is the number of
possibilities for the spin to add. In practice the execution time for
our implementation is roughly linear in $N$.


\bigskip
\bigskip

\centerline{{\bf Appendix 2: The Next\_Generation function}}
Here we give a more detailed view of this function. First, consider
the choice of $k$, the number of parents used in 
a renormalization (step 2 of the Next\_Generation
function). In our approach, $k$ is not a
fixed number, it is chosen dynamically for each renormalization in
such a way that the renormalized instance is at least $r$ times
smaller than the original one (the size of an instance is the number
of its spins). In practice, we have set $r=2.5$. We proceed
as follows. We first take a ``large'' number of 
putative parents (respecting the rule that
already used parents should not be used again unless no others are
left). For this large value of $k$, the renormalized instance
is quite big (the larger $k$, the larger the renormalized
instance). Then we decrease $k$ one unit at time until we achieve the wanted
size for the renormalized instance (or until
we reach $k=2$). Naturally, only the
$k$ finally selected parents are marked as ``used'' for the next
iteration. It may happen that the renormalized
instance is the same as the original one (with very small systems for
example). In this case, one possibility is to simply leave the
parents unchanged and use them as children (and thus directly go from
step 2 to step 6). During the recursive calls the system get smaller
and smaller; at some point, the local search is able to find the ground
state with high probability, so going to smaller sizes is useless (and
uses CPU time). To take advantage of this, we put a barrier at the size $N=15$:
$k$ is no longer decreased if there are $15$ or less spins
(and if the renormalized instance is smaller than the original one).

A second issue concerns the size of the population 
returned by Next\_Generation: we want to 
ensure that the number of
configurations in the new generation is not greater than in the old
one. As described so far, the population size could grow
because when producing the last children,
parents that have been previously used can be 
used again (as in figure~\ref{fig_next_gen}) and
thus some parents can produce more than one child. To prevent the population
from growing, 
the simplest method is to remove as many configurations as necessary
at step 8 of Next\_Generation. We have choosen to do this
by removing the worst configurations. Another possibility
would be to remove the most similar configurations (but this is
computationally more expensive). Thus step 8 is replaced by:
\begin{itemize}
\item[8'] Remove the worst configurations from the new generation so it
is not larger than the old one. Return the resulting population.
\end{itemize}

Now the Next\_Generation functions properly
but we have found it useful to introduce an algorithmic
improvement associated with preserving diversity. This is the third
point we wish to discuss. In a genetic algorithm, one evolves a
population and tries to improve it. One important characteristic
is the diversity of the population: the more the
configurations differ, the more new configurations one can create. At step
4 of Next\_Generation where the local search 
is applied, something bad can happen. Indeed, since for
small enough systems the local search is able to find the ground state
with high probability, the (local search)
optimization of $k$ different configurations
will leave us with just one
child, thereby destroying diversity. To
prevent this, we have changed step 5 as follows:
\begin{itemize}
\item[5'] If more than one configuration is left after
eliminating duplicates in step 4, proceed as in
step 5. Else raise this unique configuration and apply local
search. The children are the $k$ best configurations out of this one and
the $k$ parents.
\end{itemize}

The last point that needs to be discussed concerns how we force
Next\_Generation to return a more optimized
population than the one in its input, at least when this function
is called from 
the main program. On the contrary, when the function is called
from within the recursion,
it would be a mistake to impose strict improvement as 
the population diversity would diminish too quickly. Thus
we need to give Next\_Generation some additional information.
To do that, we add a boolean flag to its arguments
which allows one to enforce or not improvement in the new generation.
The flag is set to true when {\it initiating} the recursion, and
its value is passed onward recursively.
When the flag value is ``false'', Next\_Generation performs the 
step 5' described above. When its value is ``true'',
Next\_Generation performs instead the original
step 5. Furthermore, in the particular case 
where $k=2$ and the renormalized instance is the same as the original one, 
instead of returning both parents as children, only the best one is used. 
Finally, the flag is switched from true to false as soon as
step $4$ of Next\_Generation has improved the configurations,
and that new value is passed on recursively.

\bibliographystyle{unsrt}
\bibliography{/home/martino/Papers/Bib/references}
\end{document}